\documentclass[draftcls,onecolumn,11pt]{IEEEtran}
\usepackage{color}
\usepackage{cite}
\usepackage{epsfig}
\usepackage{amssymb}
\usepackage{amsmath}

\usepackage{graphicx}

\begin{document}
%
\title{Optimal Post-Detection Integration Technique for the Reacquisition of Weak GNSS Signals}

\author{\IEEEauthorblockN{David G\'omez-Casco, Jos\'e A. L\'opez-Salcedo and Gonzalo Seco-Granados}

\IEEEauthorblockA{IEEC-CERES, Universitat Aut\`{o}noma de Barcelona (UAB), Spain \\ Email: david.gomez.casco@uab.cat, jose.salcedo@uab.cat, gonzalo.seco@uab.cat }}

\maketitle

\begin{abstract}
This paper tackles the problem of finding the optimal non-coherent detector for  the reacquisition of weak Global Navigation Satellite System (GNSS) signals in the presence of   bits and phase uncertainty. Two  solutions  are derived  based on using two detection  frameworks: the Bayesian approach and the generalized likelihood ratio test (GLRT). We also derive approximate detectors of reduced computation complexity and without noticeable performance degradation.  Simulation results reveal a clear improvement of the detection probability for the  proposed techniques with respect to the conventional detectors implemented in high sensitivity GNSS (HS-GNSS) receivers to acquire weak GNSS signals.  Finally, we draw conclusions on which is the best technique to reacquire weak GNSS signals in practice considering a trade-off between performance and complexity. 
\end{abstract}


\section{Introduction}


Nowadays, Global Navigation Satellite System (GNSS) receivers are used in a continuously increasing variety of applications, involving for instance car and pedestrian navigation. These receivers allow the user to know the position in open-sky conditions, where the signals coming from the satellites can be easily detected.  Due to the success   achieved by GNSS receivers  in  these  conditions,   a great interest to extend their applications to more challenging environments, which include indoor building, urban canyon and forested areas, has emerged \cite{1}.


However, the existence of obstacles  in these environments causes a  high attenuation  of the received signal making the acquisition  and tracking of weak GNSS signals a challenge. In  this situation, conventional GNSS receivers are not usually able to detect the signals.  This fact has led to the development of high sensitivity GNSS (HS-GNSS) receivers. These receivers usually acquire weak signals by extending  the coherent integration time duration, which provides an additional gain in signal detection. Nevertheless, this duration  cannot be increased without boundaries mainly due to the presence of a residual frequency offset and data bits. In these circumstances, if reliable signal detection requires a longer integration time than what is possible in a coherent manner, the receiver has to apply non-linear combinations of the  coherent integration outputs, which are referred to as post-detection integration (PDI) techniques or non-coherent detectors. PDI techniques overcome the limitations of the  coherent integration time duration by using a non-linear function. Although these techniques are less effective in accumulating signal energy  than the coherent integration, they can use  a longer integration time allowing the receiver to acquire satellites with lower carrier-to-noise ratio\cite{2,2b}. 

Several PDI techniques have been proposed in  the literature  to acquire weak GNSS signals.  The best known technique is the non-coherent PDI (NPDI) \cite{3}, which is robust against the presence of frequency offset and   data bits. Another well-known option  corresponds to the differential PDI (DPDI)\cite{4}, which is only robust against the presence of frequency offset, but provides better detection performance than the  NPDI technique. An additional technique is the  generalized PDI truncated (GPDIT)  \cite{5}, which combines the two previous techniques. The GPDIT technique exhibits a gain in signal detection   with respect to the  performances of both the NPDI and DPDI techniques individually,  even though  it requires a  larger computational load and is also only robust against frequency offset.   Another alternative is the non-quadratic NPDI  (NQ-NPDI) technique,  which is robust
against data bits and residual frequency offset \cite{8250055}. This technique offers better signal detection improvement than the NPDI when the  received signal can be acquired using a small number of non-coherent combinations. Recently, a detection technique has been presented in  
\cite{6}, which is robust against the presence of  data bits but not against  frequency offset. This technique consists in the combination between two detector the NPDI and a new one referred to as squaring detector.

As a matter of fact,   although PDI techniques are usually implemented for detecting weak signals at the acquisition stage, they have  received less attention  for the reacquisition. A reacquisition  must be carried out when the receiver  has just lost the signal from  one satellite owing to,  for instance, strong attenuation caused by  an obstacle in the path between the transmitter and the receiver. If the receiver loses the signal, it  has to re-detect the signal  in order to  obtain the position of the user. However,  the problem of detecting weak GNSS signals in the reacquisition is less  complex  than  in the first acquisition  since   in case of reacquisition  an accurate estimation of the   Doppler frequency is available \cite{van2009gps} and hence the most problematic impairment to extend the coherent integration duration are the data bits. 

Despite the fact that some strategies have been proposed to  detect weak GNSS signals, which are mentioned above, the optimal PDI technique  for the reacquisition remains still unknown. This occurs because PDI techniques  are designed for the first acquisition  of the receiver,  which has to mitigate the uncertainty of the Doppler frequency.  For this reason, the purpose of this paper is to derive the optimal PDI technique by applying the detection theory tools  for the reacquisition of weak GNSS signals. More precisely,  the Bayesian approach and the generalized likelihood ratio test (GLRT) are used  to formulate the  detection problem and two PDI techniques are obtained, which require a significant amount of computational resources. We also present lower-complexity approximations of these two PDI techniques. Finally, the performance of the techniques proposed herein is compared to the PDI techniques used in previous work in terms of receiver operating characteristics (ROC) curves, revealing a clear gain in favour of our techniques.

The paper is organized as follows. Section \ref{SM} defines the signal model, while Section \ref{Sota}  makes a review of  the most relevant PDI techniques implemented in  HS-GNSS receivers. In Section \ref{DF}, new PDI techniques are derived using the Bayesian approach and the GLRT. Section \ref{Sr} illustrates the simulation results based on ROC curves. Finally, Section \ref{Cc} draws  the conclusions.



\section{ Signal Model \label{SM}}

The first task of  any GNSS receivers  is to detect  the satellites  in view. To do so,  a local replica of the transmitted  signal from a satellite with tentative values of code delay and  Doppler frequency is correlated with the  signal received from the different satellites \cite{kaplan2005understanding}. The result of this process is known as a cross-ambiguity function (CAF), which is computed for a given value of  coherent integration time ($T_{coh}$).  Assuming  that there is absence of navigation data bits transition, the  CAF of one particular satellite can be expressed as \cite{Gomez2016}
\begin{eqnarray}
 y(\tilde{\tau}, \tilde{f}_{d}) =  A d e^{j\phi} \mathrm{sinc}(\Delta f T_{coh})r(\Delta \tau)+\omega,
 \label{e_SM_1}
\end{eqnarray}
where $\tilde{\tau}$ and $\tilde{f}_{d}$ are the tentative values of code-delay and Doppler frequency, respectively, $A$ is the received amplitude  with phase $\phi$, $d$ is the data bit value that can be 1 or -1,  $\Delta \tau= \tau-\tilde{\tau}$ is the residual delay offset between  the local replica and  the received GNSS signal, $\Delta f= f_{d}-\tilde{f}_{d}$ is the  residual frequency offset, $r(\Delta \tau)$ is the  normalized correlation function of the GNSS signal, and $\omega$ is additive white Gaussian noise (AWGN) with zero-mean and variance $\sigma^2$. The $ \mathrm{sinc}(\Delta f T_{coh})$ term captures the degradation owing to the frequency offset between the local replica and the received signal.


The acquisition of a satellite provides a coarse estimation of code delay and Doppler frequency, which are obtained  from the  value of $\tilde{\tau}$ and $\tilde{f}_{d}$  that maximize  the CAF.  The accuracy of theses estimations can be improved performing a finer search of Doppler frequency and code delay in the CAF. Then, the incoming signal is tracked by correlating it with a local replica, which contains accurate estimations of Doppler frequency and code delay. This process is usually carried out for a long period of time. However, the tracking of the signal  can be lost due to,  for example,  the  attenuation  caused by an obstacle between the satellite and the receiver. In this situation, the HS-GNSS receiver  tries to reacquire the received signal from the satellite.  To do so, a local replica, which includes the  estimations of  code delay and Doppler frequency obtained in the tracking stage before losing the signal,   is correlated again with the received signal for different time instants, which becomes 
\cite{borio2009}
\begin{eqnarray}
	y_k=I_k+jQ_k=A d_k e^{j\phi}+w_k,
\label{e_SM_3}
\end{eqnarray}
where  $I_k=\Re{(y_k)}$, $Q=\Im{(y_k)}$, $w_k$ is the noise component, the  index $k =1,...,N_{nc}$  represents the time instant when the correlator output $y_k$ is computed, $d_k$ are  the data bits assumed to be a random variable taking values of 1 and -1 with the same probability. The  amplitude $A$  and the phase $\phi$  are  constant with $k$, and $w_k$ is assumed independent  for each $k$, but identically distributed. It is worth mentioning that the correlation output $y_k$  is usually computed for  several  close values of the  code delay  estimation since  this estimation may have changed slightly due to the movement of the satellite and the receiver. Nonetheless, we omit this dependence since we can consider we are performing the analysis only for one of these values.

\begin{figure}[htbp]
\begin{center}
   \vspace*{0 cm}\hspace*{0 cm}
   \epsfig{file=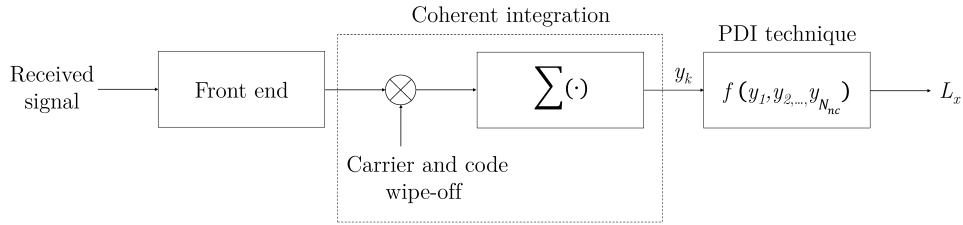,scale=0.2,angle=0}
\caption{\footnotesize{Block diagram of the GNSS  signal  reacquisition. }
\label{f_Sim_BA_0}
}
\end{center}
\end{figure}

Combinations of several correlator outputs are needed  to detect  the weak GNSS signal. The best way to obtain a gain in terms of signal detection is increasing the $T_{coh}$ (i.e. coherently combining  different correlator outputs), though its duration  is limited by  data bits.  If the coherent integration is not enough to detect the signal in harsh conditions, we must resort to apply PDI techniques, which provide  signal detection improvements since they can increase the  integration time  by using a non-linear function. In order to known whether the satellite is present or not, the  output of a PDI technique denoted as $L_{x}$ is compared  to a signal detection threshold. If  the  magnitude of $L_{x}$ surpasses the detection threshold the satellite is considered to be present, but if this magnitude does not surpass the detection threshold, the satellite is assumed to be absent. A block diagram representing the reacquisition process is shown in Fig.~\ref{f_Sim_BA_0}. The problem of obtaining the optimal  PDI technique  consists in finding a function $f(y_1,...,y_{N_{nc}})$ that allows the receiver to discriminate  between the two hypotheses $H_0$ (the satellite is absent) and $H_1$ (the satellite is present)  with the lower probability of false alarm and greater probability of detection:
\begin{itemize}
  \item Under $H_0$:  $y_k=w_k$ is a complex Gaussian noise with mean zero and variance $\sigma^2$.
  \item Under $H_1$: $y_k=Ad_ke^{j\phi}+w_k$ is the signal plus complex Gaussian noise. 
\end{itemize}

It is worth mentioning that  if the phase of the signal was time-varying, the signal detection problem would be completely different, which  leads to other types of solutions. Examples of  signal detection problems with time-varying phase  can be found in \cite{8250055,Corr2003}. 

\section{State-of-the-art of PDI techniques for HS-GNSS receivers \label{Sota}}

In this section we present a review of the most relevant PDI techniques implemented in HS-GNSS receivers, which will be used as a benchmark to compare the performance of the PDI techniques presented  in Section \ref{DF}. The optimal detector assuming a received signal that only contains an unknown phase during  all the integration time is the coherent integration \cite{richards2005} 
\begin{eqnarray}
L_{\text{coh}}(\mathbf{y})=\left|\sum_{k=1}^{N_{nc}}y_k \right|,
\label{ePDI_1}
\end{eqnarray}
where $\mathbf{y} \doteq [y_1,...,y_{N_{nc}}]^T$. However, the coherent integration is degraded, impaired in presence of  frequency offset and data bits.  In presence of these impairments, the  most widely applied PDI technique  is the NPDI, which is given by \cite{3}
 \begin{eqnarray}
L_{\text{NPDI}}(\mathbf{y})=\sum_{k=1}^{N_{nc}}\left|y_k \right|^2.
\label{ePDI_1_b}
\end{eqnarray}
The NPDI technique  is robust against the  phase variations caused by data bits and frequency offset since it removes these variations  by using the squared absolute value.  

Alternatively, another technique to detect weak signals is the DPDI defined as follows \cite{4}:
\begin{eqnarray}
L_{\text{DPDI}}(\mathbf{y})=\left|\sum_{k=2}^{N_{nc}}y_ky^*_{k-1} \right|.
\label{ePDI_1_c}
\end{eqnarray}
This technique usually offers  better performance than the NPDI technique, but it experiences a performance degradation  in presence of data bits.
Another alternative is the NQ-NPDI technique \cite{8250055}:
\begin{eqnarray}
L_{\text{NQ-NPDI}}(\mathbf{y})=\sum_{k=1}^{N_{nc}}\left|y_k \right|.
\label{ePDI_1_b2}
\end{eqnarray}
The NQ-NPDI technique provides  an improvement in signal detection  performance over the NPDI technique, especially if the signal can be detected using a  small number of $N_{nc}$, that is, $N_{nc} \le 10$. Moreover, it is robust against  frequency offset and data bits.  An additional technique, denoted as GPDIT, combines the  NPDI and DPDI techniques as \cite{5}
\begin{eqnarray}
L_{\text{GPDIT}}(\mathbf{y})=L_{\text{NPDI}}(\mathbf{y})+2L_{\text{DPDI}}(\mathbf{y})=\sum_{k=1}^{N_{nc}}|y_k|^2+2\left|\sum_{k=2}^{N_{nc}}y_ky^*_{k-1} \right|.
\label{ePDI_2}
\end{eqnarray}
The GPDIT technique outperforms the NPDI and DPDI techniques as long as  the signal does not contain data bits.  This occurs because  the GPDIT technique consists of the DPDI term, which suffers a significant degradation in presence of data bits.

\section{Detection strategies \label{DF}}

This section  uses two different detection strategies to find the optimal PDI technique for the signal model described in Section \ref{SM}. These strategies are the  Bayesian approach and the GLRT, which  are usually applied   in detection problems  with  unknown parameters. 

\subsection{Bayesian approach \label{BA}}

The Bayesian approach is   often used when likelihood ratio test (LRT) contains unknown parameters, to which a prior probability distribution can be assigned. Indeed, under these conditions, the Bayesian approach leads to the optimal detector\cite{simon1995optimum}. This approach consists in calculating the  expectation of the LRT with respect to the a priori distribution of the unknown parameter. More precisely, the difficulty caused by  the unknown parameter is circumvented by  averaging  the conditional probability density function (PDF) to obtain  the unconditional PDF, which does not depend on the unknown parameter. The conditional PDF of the correlators outputs assuming that these outputs include data bits uniformly distributed with equal probability is written  under $H_{1}$ as~\cite{8250055}
 \begin{align}
 p(\mathbf{y};H_{1}, \phi) &=   \frac{1}{(\pi\sigma^2)^{N_{nc}}}\text{exp} \left({-\sum_{k=1}^{N_{nc}} \frac{1}{\sigma^2}(I_k^2+Q_k^2+A^2)}\right)  \prod_{k=1}^{N_{nc}}\text{cosh}\left(\frac{2A}{\sigma^2}(I_k \cos(\phi)+Q_k \sin(\phi)) \right).
\label{e_BA_2}
\end{align} 
Under $H_0$, the PDF of $\mathbf{y}$  can be  expressed as follows, 
 \begin{eqnarray}
p(\mathbf{y};H_{0})=\frac{1}{(\pi\sigma^2)^{N_{nc}}}\text{exp} \left({-\sum_{k=1}^{N_{nc}} \frac{1}{\sigma^2}(I_k^2+Q_k^2)}\right).
\label{e_BA_3}
\end{eqnarray} 
 The Bayesian approach is based on a ratio of the two PDFs above given by 
\begin{align}
L_B(\mathbf{y})=\frac{\int p(\mathbf{y};H_{1}, \phi)p(\phi)d\phi}{p(\mathbf{y};H_0)}=\frac{p(\mathbf{y};H_{1})}{p(\mathbf{y};H_0)}\lessgtr \tilde{\gamma}_B,
\label{e_BA_1}
\end{align}
where $p(\phi)$ is the prior PDF of $\phi$ and $\tilde{\gamma}_B$ is the detection threshold. First,  to apply the Bayesian approach we obtain  an expression of the ratio between the two PDFs:  $p(\mathbf{y};H_{1}, \phi)$ and $p(\mathbf{y};H_0)$. After removing some irrelevant constants, the ratio can be written as
\begin{eqnarray}
L'_B(\mathbf{y},\phi) =\prod_{k=1}^{N_{nc}}\text{cosh}\left(\frac{2A}{\sigma^2}(I_k \cos(\phi)+Q_k \sin(\phi))\right ).
\label{e_BA_4}
\end{eqnarray}
 Second, we eliminate  the  phase information in (\ref{e_BA_4}) using the prior information.  The prior PDF of $\phi$  is assumed to be a uniform random variable from $-\pi$ to $\pi$. The resulting Bayesian approach is given by the following expression:
\begin{align}
L''_B(\mathbf{y})&= \frac{1}{2\pi}\int_{-\pi}^{\pi}\prod_{k=1}^{N_{nc}}\text{cosh}\left(\frac{2A}{\sigma^2}c_k(\phi)\right )d\phi 
\label{e_BA_5}
\end{align} 
 with
 \begin{eqnarray}
c_k(\phi)= I_k \cos(\phi)+Q_k \sin(\phi).
\label{e_BA_6}
\end{eqnarray} 
Note that  the larger the value of  $N_{nc}$, the larger the number of  multiplicative terms in the integral. To solve this integral, we apply the properties of the product of cosh functions, that is, $\text{cosh}(x)\text{cosh}(y)=(\text{cosh}(x+y)+\text{cosh}(x-y))/2$. Proceeding in this way the integral can be rewritten as a series of integrals, where each one contains the cosh of a certain combination of sums and subtractions of the terms $c_k(\phi)$ as
\begin{align}
L''_B(\mathbf{y}) &= \frac{1}{2\pi 2^{N_{nc}-1}}\left(\int_{-\pi}^{\pi}\text{cosh}\left(\frac{2A}{\sigma^2}(c_1(\phi)+c_2(\phi)+\hdots+c_{N_{nc}}(\phi))\right) d\phi  \right. +\hdots+ \nonumber \\  & \left. \int_{-\pi}^{\pi}\text{cosh}\left(\frac{2A}{\sigma^2}(c_1(\phi)-c_2(\phi)-\hdots-c_{N_{nc}}(\phi))\right) d\phi \right),
\label{e_BA_7}
\end{align}
for which a more compact expression is
\begin{eqnarray}
L''_B(\mathbf{y}) = \frac{1}{2\pi 2^{N_{nc}-1}}\sum_{m=1}^{2^{N_{nc}-1}} \int_{-\pi}^{\pi}\text{cosh}\left(\frac{2A}{\sigma^2}(a_m\cos(\phi)+b_m\sin(\phi))\right) d\phi,  
\label{e_BA_8}
\end{eqnarray}
where $2^{N_{nc}-1}$ is the number of cosh functions that appear after applying the property of the multiplication of several cosh functions. The  $a_m$ and $b_m$ coefficients aim at encompassing all possible combinations of additions and subtractions of $I_k$ and $Q_k$, respectively, excluding those that refer to others already considered but with opposite sign.  By stacking the above-mentioned coefficients into vectors $\mathbf{a}\doteq[a_1,\cdots,a_{2N_{nc}-1}]^T$ and $\mathbf{b}~\doteq~[b_1,\cdots,b_{2N_{nc}-1}]^T$, we can compute their value as follows,
\begin{align}
\label{e_BA_8_b}
\mathbf{a}&=\mathbf{MI} \\
\mathbf{b}&=\mathbf{MQ},
\label{e_BA_8_c}
\end{align}
where $\mathbf{I}\doteq[I_1,\hdots,I_{N_{nc}}]^T$, $\mathbf{Q}\doteq[Q_1,\hdots,Q_{N_{nc}}]^T$, and $\mathbf{M}$ is a $\left(2^{N_{nc}-1}\times N_{nc}\right)$  matrix whose rows contain all the possible combinations of +1 and -1, excluding those that differ from another row in a global change of sign as
 \begin{eqnarray} 
\mathbf{M}\doteq
  \begin{bmatrix}
    1 & -1 & -1 & \ldots & -1 \\
     1 & 1 & -1 & \ldots & -1 \\
      \vdots &  \vdots &  \vdots & \ddots &  \vdots \\
      1 & 1 & 1 & \ldots & 1 \\
  \end{bmatrix}.
 \label{e_BA_11}
\end{eqnarray}

 Now, the integral can be solved by the following procedure as
\begin{align}
L''_B(\mathbf{y}) &= \frac{1}{\pi 2^{N_{nc}}}\sum_{m=1}^{2^{N_{nc}-1}} \int_{-\pi}^{\pi}\text{cosh}\left(\frac{2A}{\sigma^2}\sqrt{a_m^2+b_m^2}\cos\left(\phi-\text{atan}\left(\frac{b_m}{a_m}\right) \right)\right) d\phi  \nonumber \\ &= \frac{1}{\pi 2^{N_{nc}+1}}\sum_{m=1}^{2^{N_{nc}-1}} \left(\int_{-\pi}^{\pi}e^{\frac{2A}{\sigma^2}\sqrt{a_m^2+b_m^2}\cos(\phi)}d\phi+\int_{-\pi}^{\pi} e^{-\frac{2A}{\sigma^2}\sqrt{a_m^2+b_m^2}\cos(\phi)}d\phi\right)\nonumber \\ &= \frac{1}{ 2^{N_{nc}-1}}\sum_{m=1}^{2^{N_{nc}-1}}I_0\left( \frac{2A}{\sigma^2}\sqrt{a_m^2+b_m^2}\right),
\label{e_BA_10}
\end{align}
 where $I_0$ denotes the zero-order modified Bessel function. Finally, removing some irrelevant constants  the resulting detector can be expressed as
\begin{align} 
L_\text{BAPDI}(\mathbf{y})=\sum_{m=1}^{2^{N_{nc}-1}}I_0\left( \frac{2A}{\sigma^2}\sqrt{a_m^2+b_m^2}\right)\lessgtr \gamma_B,
\label{e_BA_10_B}
\end{align} 
where $\gamma_B$  is the detection threshold. The result expressed in (\ref{e_BA_10_B}) is referred to as  Bayesian approach PDI (BAPDI) technique. This technique  is optimum  in presence of unknowns bits and an unknown constant phase. Nonetheless, the BAPDI technique depends on the ratio of   $A$ and $\sigma^2$. Despite the fact that some receivers can know this ratio in tracking stage since they use a carrier-to-noise estimator, the goal of this paper  is to derive a detector which does not depends on the parameters $A$ and $\sigma^2$ so that it can be implemented in any receiver. To do so, we propose to apply the approximation of $I_0(x) \approx \text{exp}(|x|)$, valid for relative large values of $x$. This approximation  can be applied for our problem since 
the argument of (\ref{e_BA_10_B}) is not a small magnitude when the received signal has the same or similar combination  of bits as one of the rows of the matrix $\mathbf{M}$.   Then,  by considering  $I_0(x) \approx \text{exp}(|x|)$,  we get
\begin{eqnarray}
 \sum_{m=1}^{2^{N_{nc}-1}}\text{exp}\left( \left| \frac{2A}{\sigma^2}\sqrt{a_m^2+b_m^2}\right|\right)\lessgtr \gamma_B.
 \label{e_BA_12}
\end{eqnarray} 
 Introducing the logarithm of the LRT becomes
\begin{eqnarray}
  \ln\left( \sum_{m=1}^{2^{N_{nc}-1}}\text{exp}\left( \left| \frac{2A}{\sigma^2}\sqrt{a_m^2+b_m^2}\right|\right)\right)\lessgtr \ln(\gamma_B).
 \label{e_BA_13}
\end{eqnarray}
To simplify the expression above, we make use of the log-sum-exp approximation, which consists in taking the maximum of the different exponentials. This approximation is  reasonable for a  high signal-to-noise ratio (SNR) at the output of the PDI technique. Such values of SNR are usually obtained at this output because otherwise the signal could not be detected. In this situation,  the largest term dominates  in the sum of (\ref{e_BA_13})  as   
\begin{eqnarray}
 \max_m\left(  \left| \frac{2A}{\sigma^2}\sqrt{a_m^2+b_m^2}\right|\right)\lessgtr \gamma_B'.
 \label{e_BA_14}
\end{eqnarray}
The larger the deviation of the argument of (\ref{e_BA_13}), the better the approximation becomes.   Finally, incorporating the now irrelevant constant $\frac{2|A|}{\sigma^2}$  into the threshold,  the resulting detector can be  expressed as
\begin{eqnarray}
L_\text{MBAPDI}(\mathbf{y})=\max_m\left(  \sqrt{a_m^2+b_m^2}\right)\lessgtr \gamma_B''.
 \label{e_BA_15}
\end{eqnarray}
The solution provided by (\ref{e_BA_15}) is referred to as maximum BAPDI (MBAPDI) technique.  The MBAPDI  technique can be implemented in any HS-GNSS receiver because it does not depend on the parameters $A$ and $\sigma^2$. It is worth mentioning that if chosen index $m$ corresponds to the correct sequence of bits, then the result would be the same as for the  coherent detector in the hypothesis $H_1$, but this will not always happen due to the presence of noise. Moreover, although this happened,  we would have  a performance degradation  with respect to the coherent detector. This is because  the MBAPDI requires the use of the maximum  function also in the hypothesis $H_0$, making the receiver  choose the largest value among the different $2^{N_{nc}-1}$ samples of noise, which  increases the number of false alarms.

\subsection{Generalized likelihood ratio test}

A common approach to design detectors with unknown parameters deals with the combination of estimation and detection. The best known joint estimation and detection approach is the GLRT, which consists of two steps. First, the maximum likelihood (ML) estimate of the unknown parameters are found. Second, the unknown parameters are replaced by their ML estimates under each hypothesis and  the LRT is calculated as if the estimated parameters were correct \cite{DoSiN,levy_det}.

Although no claims about the optimality of the GLRT can be made, it  provides good results in general. Moreover, the GLRT formulation usually provides simpler expressions than the Bayesian approach, which  requires the integral of products of several PDFs. This occurs because ML estimation equations  sometimes result in a closed-form solution. However, this is not the case of our problem where the ML estimate of the received phase  affected by bits does not admit a closed-form solution. In this situation, two options are feasible: making an approximation of the ML equation in order to get a closed-form solution, which was done in \cite{borio2009} or using a one-dimensional search method to evaluate the ML estimate.

A PDI technique has been already obtained in \cite{6} using an approximation of the ML phase estimate provided in \cite{borio2009}   and replacing it in an expression of the LRT approximated for low SNR regime.  
Before proceeding, we make a brief description of the work done previously by others authors and after that we present new   PDI techniques  based on  using different approaches of the GLRT. In \cite{borio2009}, the authors computed the ML solution of the signal phase, which contains unknown bits, from the PDF of $\mathbf{y}$ as
\begin{align}
 p(\mathbf{y};H_{1}, \phi) &=   \frac{1}{(\pi\sigma^2)^{N_{nc}}}\text{exp} \left({-\sum_{k=1}^{N_{nc}} \frac{1}{\sigma^2}(I_k^2+Q_k^2+A^2)}\right)  \prod_{k=1}^{N_{nc}}\text{cosh}\left(\frac{2A}{\sigma^2}(I_k \cos(\phi)+Q_k \sin(\phi)) \right).
\label{e_GL_1}
\end{align}
 The log-likelihood function for $\phi$,  removing the terms that are not affected by $\phi$, can be expressed as 
\begin{eqnarray}
 L(\mathbf{y},\phi) =     \sum_{k=1}^{N_{nc}}\text{ln}\left(\text{cosh}\left(\frac{2A}{\sigma^2}(I_k \cos(\phi)+Q_k \sin(\phi)) \right)\right).
\label{e_GL_2}
\end{eqnarray}
 In order to find a closed-form solution of  $\phi$ the $\ln(\text{cosh}(x))$ function is approximated by $x^2/2$. Thus, the closed-form expression of $\phi$ that approximately maximizes (\ref{e_GL_2})     is  
\begin{eqnarray}
\hat{\phi}=\frac{1}{2}\text{atan2}\left(2\sum_{k=1}^{N_{nc}}I_kQ_k,\sum_{k=1}^{N_{nc}}I_k^2-Q_k^2 \right),
\label{e_GL_3}
\end{eqnarray}
where $\text{atan2}(x,y)$ is the four quadrant atan function.

\begin{figure}[htbp]
\begin{center}
   \vspace*{0 cm}\hspace*{0 cm}
   \epsfig{file=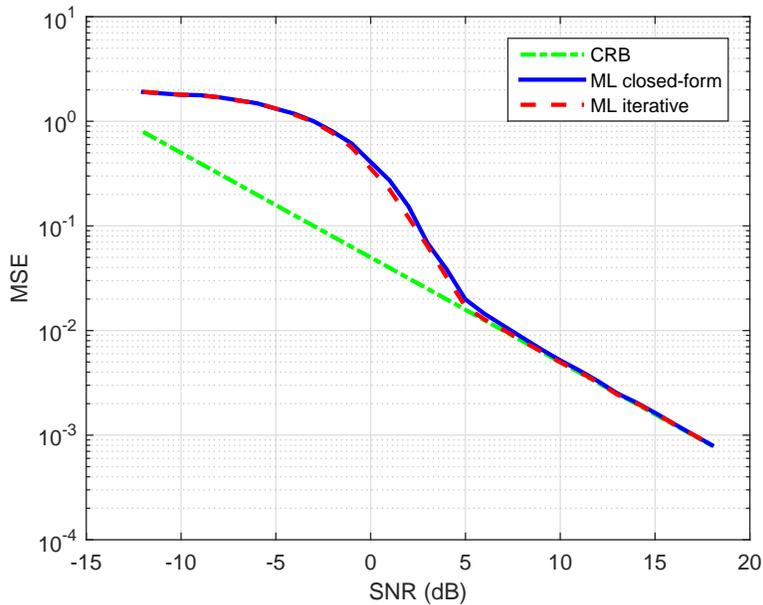,scale=0.8,angle=0}
\caption{\footnotesize{Performance  comparison of the different  estimators and the CRB for $N_{nc}=10$. The approximation of the ML solution  expressed in (\ref{e_GL_3}) is referred to as  ML closed-form  and the one obtained from  a one-dimensional search is indicate as ML iterative.}
\label{f_GL_1}
}
\end{center}
\end{figure}

Another way to find the  value of $\phi$  that maximizes  (\ref{e_GL_2})  is by using an iterative algorithm. It can be easily carried out implementing a one-dimensional search. The comparison  between the estimators and the Cramer-Rao bound (CRB) is shown in Fig.\ref{f_GL_1}. The result illustrates that the ML estimate obtained by a one-dimensional search method exhibits practically the same performance as the approximation in (\ref{e_GL_3}).  The CRB of the phase estimate is    $1/(2\text{SNR}N_{nc})$ \cite{CRB_f}, where the SNR is defined as $A^2/\sigma^2$.

In \cite{6}, a PDI technique was presented based on the GLRT approach. The authors  used the log-LRT, which can be expressed as (\ref{e_GL_2}). They propose to   approximate $L(\mathbf{y},\phi)$ defined in (\ref{e_GL_2}) for  a low SNR regime applying a Taylor series of the $\text{ln}(\text{cosh}(x))$ function as $x^2/2$, which leads to 
\begin{eqnarray}
\sum_{k=1}^{N_{nc}} \left(\frac{2A}{\sigma^2}(I_k \cos(\phi)+Q_k \sin(\phi))\right )^2 \lessgtr \gamma'_G.
\label{e_GL_5}
\end{eqnarray}
Replacing  the  approximation of the phase estimate in (\ref{e_GL_3}) into (\ref{e_GL_5}), and making some simplifications, the NPDISD detector  can be obtained as
\begin{eqnarray}
L_\text{NPDISD}(\mathbf{y}) =\sum_{k=1}^{N_{nc}} |y_k|^2+\left|\sum_{k=1}^{N_{nc}} y_k^2\right|.
\label{e_GL_6}
\end{eqnarray}
The NPDISD detector consists  of two non-coherent detectors or PDI techniques.  The first detector  is the  conventional NPDI detector. The second detector is the squaring detector (SD), which consists in summing the squared complex correlator outputs. Despite the fact that this solution provides a good performance, an enhancement of this approach can be carried out since HS-GNSS receivers do not usually work in a very low SNR regime at the output of the coherent correlation. This occurs because the correlator outputs in HS-GNSS receivers are obtained using a long $T_{coh}$ in general and combining few correlation outputs the signal can be detected. Then, the approximation of Taylor series used in (\ref{e_GL_5}) for low SNR values might not be the best option to obtain the best performance  of the GLRT approach.

For this reason, the purpose of the following subsections is to propose several new alternatives to the GLRT in order to enhance the  performance of the NPDISD technique. More precisely,  we present three new  approaches  to obtain  the best detectors using the GLRT approach in the context of HS-GNSS receivers. 
 
 \subsubsection{GLRT (strict)\label{GLRT_stric}}
 
The first one boils down  to the  strict application of the GLRT approach.  This approach is based on using  the log-LRT  and replacing  the unknown parameter $\phi$  value by its ML estimation, which must be obtained from a one-dimensional search in  (\ref{e_GL_2}), as
\begin{eqnarray}
L(\mathbf{y},\hat{\phi}_\text{ML}) =\sum_{k=1}^{N_{nc}}\ln\left(\text{cosh} \left(\frac{2A}{\sigma^2}(I_k \cos(\hat{\phi}_\text{ML})+Q_k \sin(\hat{\phi}_\text{ML}))\right ) \right ),
\label{e_GL_7}
\end{eqnarray}
where $\hat{\phi}_\text{ML}$ is the ML estimate of $\phi$. This approach allow us to know, which is the optimal performance of the GLRT method and how far  it is from the Bayesian approach. This is an important point because  the outcome of the Bayesian approach is the optimal detector under  the assumed conditions. As we have seen  in  Subsection \ref{BA}, the result of the Bayesian approach implies the computation of a matrix, whose size increases exponentially  with the $N_{nc}$  value. In fact, the computation of this matrix  can become a handicap.  For this reason, if the difference between  the performance of the Bayesian approach and the GLRT was  quite similar, the application of the GLRT could be the best option. We will continue this discussion  later on in the Section \ref{Sr}  where the performance comparison of the PDI techniques is analysed.

\subsubsection{GLRT approximation in closed-form\label{GLRT_a}} 

The second approach is based on the log-LRT in (\ref{e_GL_7}), but reducing the complexity of this method  to estimate $\phi$.  Given the phase estimate in (\ref{e_GL_3}) exhibits almost the same performance as at ML phase estimate, while avoiding the one dimensional search, we propose to replace $\hat{\phi}_\text{ML}$ in (\ref{e_GL_7}) with (\ref{e_GL_3}), resulting in:
\begin{eqnarray}
L(\mathbf{y},\hat{\phi}) =\sum_{k=1}^{N_{nc}}\ln\left(\text{cosh} \left(\frac{2A}{\sigma^2}(I_k \cos(\hat{\phi})+Q_k \sin(\hat{\phi}))\right ) \right ).
\label{e_GL_8}
\end{eqnarray}

\subsubsection{GLRT approximation for  high SNR regime\label{GLRT_a_l}} 
 
  The alternatives described  in Subsection  \ref{GLRT_a} and     Subsection \ref{GLRT_stric} require the knowledge of the SNR, $A/\sigma^2$. This is a drawback  since this information is sometimes   unknown by the receiver. For this reason, the last method that we propose avoids the need of knowing  the SNR a priori.  The way  to  obtain a detector that does not depend on the SNR is to adopt an approximation of the $\ln(\text{cosh}(x))$ function as $|x|-\ln(2)$. This approximation gives an excellent fit  for  relative large values of $x$, which is a region  appropriate to detect signals in the context of HS-GNSS receivers. After using this approximation, the PDI technique  is independent of the scale factors $A$ and $\sigma^2$. Thus, the resulting technique can be expressed  as 
  \begin{eqnarray}
L_\text{GLRT$_{a. l.}$}(\mathbf{y},\hat{\phi}) =\sum_{k=1}^{N_{nc}} \left|I_k \cos(\hat{\phi})+Q_k \sin(\hat{\phi}) \right|.
\label{e_GL_9}
\end{eqnarray} 
This  PDI technique may offer   similar performance as the two previous techniques  presented  in  Subsection \ref{GLRT_stric} and Subsection \ref{GLRT_a} when the SNR at the  correlator output is relatively high. Besides not requiring the knowledge of the SNR, this technique avoids the use of two non-linear functions such as the $\ln$ and $\cosh$. It is worth mentioning that (\ref{e_GL_9}) has some resemblance to the  NQ-NPDI technique described in (\ref{ePDI_1_b2}), which was derived  for time-varying phase signals. The NQ-NPDI technique  offers a  great performance in scenarios where the SNR is relatively high and the received signal  can suffer  phase changes \cite{8250055}. However,  the technique proposed in this subsection is derived   for signals with constant phase. This fact suggests that in scenarios where the  received signal  includes a  constant phase, the detector in  (\ref{e_GL_9}) could provide a promising performance.

\section{Simulation results \label{Sr}}

This section presents the simulation results based on  receiver operating characteristic (ROC) curves.   These curves compare  the detection performance of the PDI techniques  proposed  herein   to  the more relevant  PDI techniques found in the literature.  Results are obtained  using  Monte Carlo simulations and the $\sigma$ value is normalized to 1.

\begin{figure}[tbp]
\begin{center}
   \vspace*{0 cm}\hspace*{0 cm}
   \epsfig{file=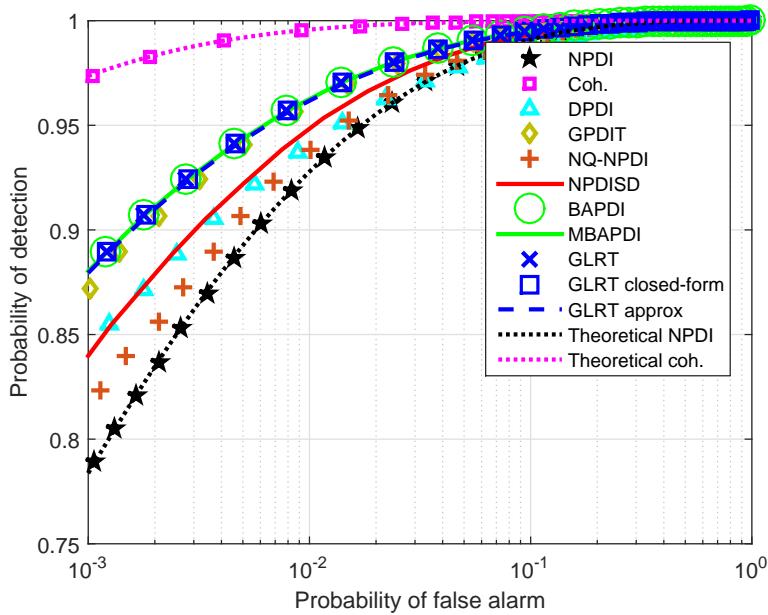,scale=0.8,angle=0}
\caption{\footnotesize{Performance comparison of the different detectors in absence of bits for $N_{nc}=6$, $A=1.6$  and $\sigma=1$. In the legend, GLRT, GLRT closed-form and GLRT approx refer to the techniques explained in Subsections \ref{GLRT_stric}, \ref{GLRT_a} and \ref{GLRT_a_l}, respectively. }
\label{f_Sim_BA_1}
}
\end{center}
\end{figure}

Fig. \ref{f_Sim_BA_1}  shows the comparison among the different PDI techniques in an ideal channel containing only Gaussian noise and  an unknown constant phase, but in absence of  data bits  in the received signal for $N_{nc}=6$ and $A=1.6$. As we expected, in this situation, the optimal detector is the coherent integration since there are not effects that pose limits on its duration. The worst performing technique corresponds to the NPDI technique. The proposed five techniques, namely, BAPDI, MBAPDI, and the three obtained from the  GLRT method explained in Subsections \ref{GLRT_stric}, \ref{GLRT_a} and \ref{GLRT_a_l}, exhibit similar performance, which is also better than that of the DPDI, GPDIT, NPDISD and NQ-NPDI techniques. Theoretical ROC curves of the  coherent integration and the NPDI  technique are included to the figures, which are given by 
\cite{richards2005} and \cite{jayaram2013noncoherent}, respectively. However, for the rest of the detectors there are no known closed-form expressions of their ROC curves.

\begin{figure}[tbp]
\begin{center}
   \vspace*{0 cm}\hspace*{0 cm}
   \epsfig{file=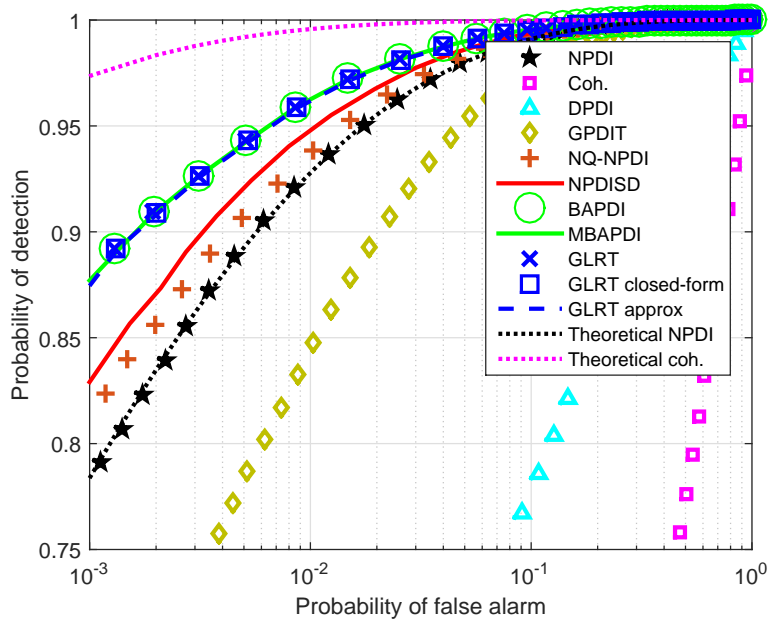,scale=0.8,angle=0}
\caption{\footnotesize{Performance comparison of the different detectors in presence of data bits for $N_{nc}=6$, $A=1.6$ and $\sigma=1$. In the legend, GLRT, GLRT closed-form and GLRT approx refer to the techniques explained in Subsections \ref{GLRT_stric}, \ref{GLRT_a} and \ref{GLRT_a_l}, respectively.}
\label{f_Sim_BA_2}
}
\end{center}
\end{figure}

Fig.   \ref{f_Sim_BA_2} shows the comparison among the different detectors in a Gaussian channel when the received signal is affected by phase changes owing to  data bits using the same parameters as in Fig. \ref{f_Sim_BA_1}.  The result illustrates that the DPDI, GPDIT and the coherent integration techniques suffer a strong degradation performance since they are not robust  against the presence of bits. In this case, the proposed five techniques, two based on the Bayesian approach and three established from the GLRT, provide a very similar performance  outperforming  the rest of the PDI techniques. In particular, it is interesting to  pay attention to  the comparison between the proposed five  techniques and the NPDISD technique, which was derived by the application of the GLRT approach, but the author used an approximation  for a low SNR regime. The outcome reveals a clear improvement in favour of the techniques proposed herein.

\begin{figure}[htbp]
\begin{center}
   \vspace*{0 cm}\hspace*{0 cm}
   \epsfig{file=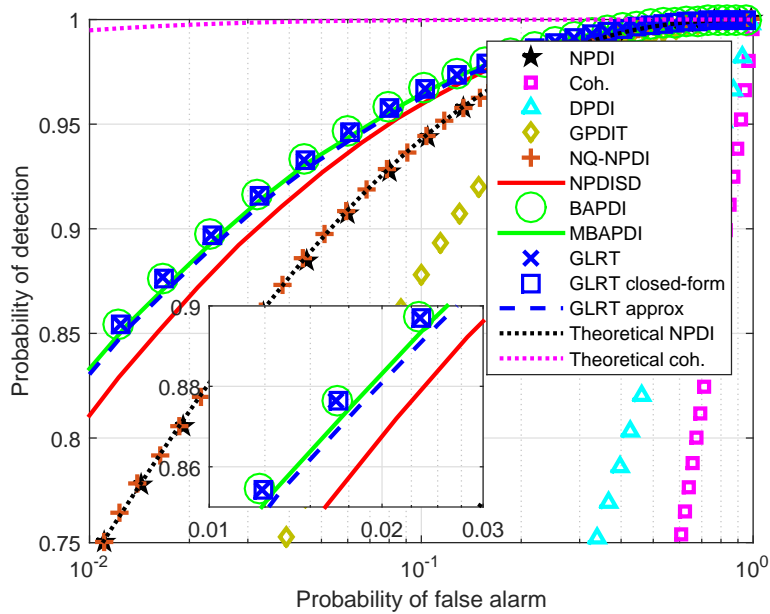,scale=0.8,angle=0}
\caption{\footnotesize{Performance comparison of the different detectors in presence of data bits for $N_{nc}=15$, $A=1$ and $\sigma=1$. In the legend, GLRT, GLRT closed-form and GLRT approx refer to the techniques explained in Subsections \ref{GLRT_stric}, \ref{GLRT_a} and \ref{GLRT_a_l}, respectively.}
\label{f_Sim_BA_3}
}
\end{center}
\end{figure}

 Fig. \ref{f_Sim_BA_3} illustrates  the comparison among the different detectors in a Gaussian channel when the received signal contains unknown  data bits  for $A=1$ and $N_{nc}=15$. This simulation reveals that although the SNR of the correlator output is lower than in Fig.   \ref{f_Sim_BA_1} and Fig.   \ref{f_Sim_BA_2}, the proposed five techniques remain exhibiting the best performances.  The performance difference among the five techniques and the NPDISD is smaller  than in the previous simulations due to this lower SNR value.  This  value also causes that the technique described in  Subsection \ref{GLRT_a_l}, which has been derived for a relative large values of SNR, has a slight mismatch with respect to the techniques defined in Subsection \ref{GLRT_stric} and  Subsection  \ref{GLRT_a}. The MBAPDI technique also offers a very slight degradation with respect to the BAPDI because the SNR at the output of the PDI technique is slightly lower than in  Fig. \ref{f_Sim_BA_2}. This effect can be seen  in the zoom view, which appears in Fig. \ref{f_Sim_BA_3}.

\begin{figure}[tbp]
\begin{center}
   \vspace*{0 cm}\hspace*{0 cm}
   \epsfig{file=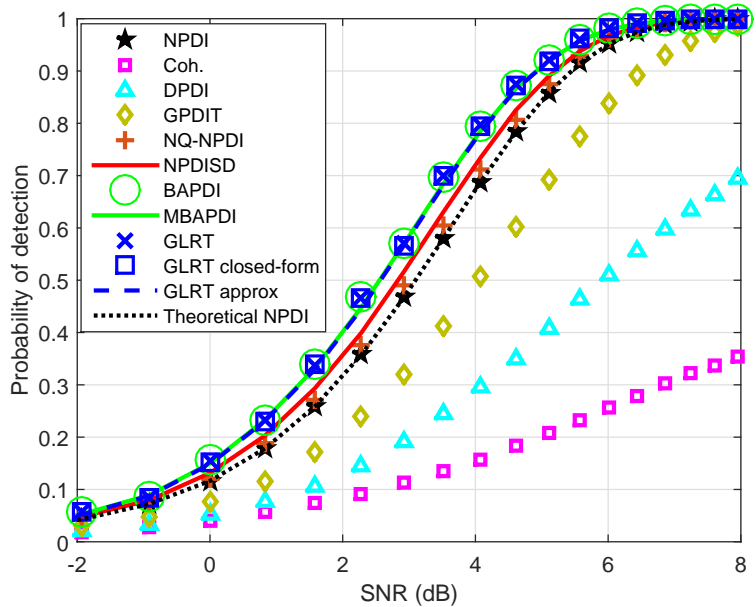,scale=0.8,angle=0}
\caption{\footnotesize{Probability of detection vs. SNR with $N_{nc}=5$ and  probability of false alarm of 1e-3 for the different detectors  in presence of data bits in the received signal.}
\label{f_Sim_BA_4}
}
\end{center}
\end{figure}

Fig. \ref{f_Sim_BA_4} shows the probability of detection with respect to the SNR for the different detectors   in a Gaussian channel and when the received signal contains data bits. We use $N_{nc}=5$ and set the probability of false alarm to $1e-3$. The detection threshold for each PDI technique is fixed  through the Monte Carlo simulations. The result illustrates that the techniques proposed in the paper show the highest probabilities of detection.  The coherent integration, DPDI and  GPDIT techniques suffer  a severe degradation   due to the  data bits. For this reason,  these techniques   are not useful in detection problems where the received signal have sign changes produced by the bits. The NPDI, NQ-NPDI and NPDISD techniques, which are robust against the presence of data bits, outperform the coherent integration, DPDI and  GPDIT techniques, but the former group does not provided as good performance  as the techniques  presented in this work. 

Given that the five techniques presented in the paper offer very similar performance, exceeding that of the other techniques, for the problem at hand, the selection of the most suitable one can be based on the computational complexity. While the BAPDI is the theoretically optimal PDI technique since it has been derived from the Bayesian approach, it may present difficulties in practice because it uses a matrix, whose size grows exponentially as $N_{nc}$ grows. Moreover, the BAPDI requires the a priori knowledge of the SNR  and it needs to use the modified Bessel function, which in practice has to be evaluated numerically. The  MBAPDI  also suffers the disadvantage of  having to evaluate a potentially large number of combinations,  which introduces a large computational burden, especially for large $N_{nc}$ values.  The  exact GLRT   presented in Subsection \ref{GLRT_stric}  requires the usage of a one-dimensional search method  to estimate the phase of the received signal. This fact poses difficulties in the implementation of this technique in a HS-GNSS receiver. The GLRT approach described in Subsection \ref{GLRT_a} is a good option since it does not depend on large matrices nor  a one-dimensional search method, but it has the drawback of requiring the knowledge of the SNR. Finally,  the PDI technique presented in Subsection \ref{GLRT_a_l} becomes the best  option to obtain a  significant gain in terms of signal detection because its computational load is the lowest one and it does not need a priori information about the SNR.
   

\section{Conclusions \label{Cc}}

In this paper we have derived  two PDI techniques by using the Bayesian approach and  the GLRT for the reacquisition of weak GNSS signals.  We have also proposed  approximate   techniques of reduced computational complexity, which can be easily implemented  in  HS-GNSS receivers and  do not require the knowledge of the SNR.  Simulation results have shown the superior performance of the techniques proposed in the paper with respect to other PDI techniques, while  the former group   provides very similar performance.  For  a balanced trade-off between  computational burden and performance, we can conclude that the  most suitable technique for the reacquisition of GNSS signals is the one  based on the approximation of the GLRT  approach  for high SNR regime and on the use of the approximate ML phase estimate.  

\bibliographystyle{IEEEtran}

\bibliography{strings}

\end{document}